\documentstyle[12pt]{article}

\newcommand{\eq}{\begin{equation}}
\newcommand{\eqa}{\begin{eqnarray}}
\newcommand{\en}{\end{equation}}
\newcommand{\ena}{\end{eqnarray}}
\newcommand{\enn}{\nonumber \end{equation}}

\def\sk{\vskip .4cm}
\def\noi{\noindent}
\def\om{\omega}

\def\la{q-q^{-1}}
\def\linv{{1 \over \la}}
\def\lam{{1 \over \la}}

\def\ab{\bar{a}}
\def\Ab{\bar{A}}
\def\Bb{\bar{B}}
\def\Cb{\bar{C}}
\def\Db{\bar{D}}
\def\ab{\bar{a}}

\def\bb{\bar{b}}

\def\ub{\bar u}
\def\vb{\bar v}
\def\xib{\bar \xi}

\def\epsi{\varepsilon}
\def\we{\wedge}

\def\de{\delta}

\def\part{\partial}

\def\A#1#2{ A^{#1}_{~~~#2} }

\def\R#1#2{ R^{#1}_{~~~#2} }
\def\Rp#1#2{ (R^+)^{#1}_{~~~#2} }
\def\Rpinv#1#2{ [(R^+)^{-1}]^{#1}_{~~~#2} }

\def\Rinv#1#2{ (R^{-1})^{#1}_{~~~#2} }

\def\Rbo{{\cal R}}
\def\Rb#1#2{{ \Rbo^{#1}_{~~~#2} }}

\def\Rbpm#1#2{(\Rbo^{\pm})^{#1}_{~~~#2} }

\def\Rh{{\hat R}}

\def\Rhat#1#2{ \Rh^{#1}_{~~~#2} }
\def\L#1#2{ \La^{#1}_{~~~#2} }

\def\La{\Lambda}

\def\ff#1#2#3{f_{#1~~~#3}^{~#2}}

\def\cchi#1#2{\chi^{#1}_{~#2}}
\def\ome#1#2{\om_{#1}^{~#2}}
\def\RRhat#1#2#3#4#5#6#7#8{\La^{~#2~#4}_{#1~#3}|^{#5~#7}_{~#6~#8}}

\def\LL#1#2#3#4#5#6#7#8{\La^{~#2~#4}_{#1~#3}|^{#5~#7}_{~#6~#8}}

\def\Cb{{\bf C}}
\def\CC#1#2#3#4#5#6{\Cb_{~#2~#4}^{#1~#3}|_{#5}^{~#6}}

\def\C#1#2{ {\bf C}_{#1}^{~~~#2} }

\def\Dmat#1#2{D^{#1}_{~#2}}
\def\Dmatinv#1#2{(D^{-1})^{#1}_{~#2}}

\def\T#1#2{ T^{#1}_{~~#2} }
\def\Ti#1#2{ (T^{-1})^{#1}_{~~#2} }

\def\Th#1#2{ {\hat T}^{#1}_{~~#2} }

\def\qm{q^{-1}}

\def\D{\Delta}

\def\Dp{\Delta^{\prime}}
\def\Ip{I^{\prime}}
\def\ep{\epsi^{\prime}}
\def\kp{\kappa^{\prime}}

\def\qone{q \rightarrow 1}

\def\Lpm#1#2{(L^{\pm})^{#1}_{~~#2}}
\def\Lmp#1#2{(L^{\mp})^{#1}_{~~#2}}

\def\Lp#1#2{(L^{+})^{#1}_{~~#2}}
\def\Lm#1#2{(L^{-})^{#1}_{~~#2}}

\def\detq{{\det}_q}

\begin{document}

\begin{titlepage}
\rightline{DFTT-57/92}
\rightline{October 1992}
\vskip 2em
\begin{center}{\bf ON THE QUANTUM POINCAR\'E GROUP }
\\[6em]
 Leonardo Castellani\\[2em]
{\sl Istituto Nazionale di
Fisica Nucleare, Sezione di Torino\\
and\\Dipartimento di Fisica Teorica\\
Via P. Giuria 1, 10125 Torino, Italy.}  \\[6em]
\end{center}
\begin{abstract}
The inhomogeneous quantum groups $IGL_q(n)$ are obtained by means of a
particular projection of $GL_q(n+1)$. The bicovariant differential
calculus on $GL_q(n)$ is likewise projected into a consistent
bicovariant calculus on $IGL_q(n)$. Applying the same method to $GL_q(n,
\Cb)$ leads to a bicovariant calculus for the complex
inhomogeneous quantum groups $IGL_q(n,\Cb)$.
The quantum
Poincar\'e group and its bicovariant geometry are recovered by
specializing our results to $ISL_q(2,\Cb)$.
\end{abstract}

\vskip 4cm

\noi DFTT-57/92

\noi October 1992
\vskip .2cm
\noi \hrule
\vskip.2cm
\hbox{\vbox{\hbox{{\small{\it email addresses:}}}\hbox{}}
 \vbox{\hbox{{\small Decnet=31890::CASTELLANI}}
\hbox{{\small Bitnet= CASTELLANI@TO.INFN.IT }}}}

\end{titlepage}
\newpage
\setcounter{page}{1}


The study of a continuous deformation of the Poincar\'e
group is worthwhile
{\sl per se}, given the central
role of this group in physics. In the context of
a quantum group theoretic formulation of gravity
theories, it is in fact essential to
find a consistent $q$-deformation of
the Poincar\'e group. This we will present in this
letter.
\sk
Quantum groups \cite{Drinfeld}-\cite{Majid1} have
emerged in the last years as nontrivial
deformations of Lie groups, and the
differential calculus on them has been developed
recently \cite{Wor,Bernard,Jurco,Zumino,Aschieri}. The
general constructive procedure of ref. \cite{Jurco} works
for the $q$-groups of the $A,B,C,D$ series, and in ref. \cite{Cas1} we
have studied how to
extend it to
nonhomogeneous quantum groups.
\sk
We begin by presenting a general method for constructing inhomogeneous
quantum groups and their complexification. In ref. \cite{Cas1}
we have found the $R$-matrix and a bicovariant differential
calculus for $IGL_q(n)$, using the definition of inhomogeneous
$q$-groups as given in \cite{Schlieker}. Here, however, we take a
different route and obtain the inhomogeneous $q$-groups $IGL_q(n)$
and $IGL_q(n,\Cb)$ (and their bicovariant \footnote{``bicovariant"
meaning that one can define a left and a right action of the $q$-group
on the space of quantum one-forms, as in the $q=1$ case,
see ref. \cite{Wor}.} differential calculi)
as {\sl projections} of $GL_q(n+1)$ and $GL_q(n+1,\Cb)$ (and their
bicovariant differential calculi).
Both procedures are equivalent and lead to the same $q$-differential
calculi; the one we present here has the advantage of giving
a new interpretation to the results of ref.s \cite{Schlieker,Cas1}.
\sk
Finally, our method is specialized to $ISL_q(2,\Cb)$,
the quantum Poincar\'e group, whose bicovariant $q$-Lie algebra
is found and given
explicitly in the Table.
\sk
Other papers concerning the quantum Lorentz group or the quantum
Poincar\'e group are quoted in
\cite{Drabant}-\cite{qPoincare}.
\sk
The key observation is that the $R$-matrix of $GL_q(n+1)$ can be
written as ({\small A=(0,a)}):
\eq
\R{AB}{CD}=\left( \begin{array}{cccc}
                  q&0&0&0\\0&1&0&0\\0&q-\qm&1&0\\
                  0&0&0&\R{ab}{cd}\\
                \end{array} \right) \label{RGL}
\en
\noi where $\R{ab}{cd}$ is the $R$-matrix of $GL_q(n)$, and
the indices {\small AB} are
ordered as {\small $00,0b,a0,ab$}. This form of the $R$-matrix
allows a consistent projection of $GL_q(n+1)$ into $GL_q(n)$.
This projection works also for the corresponding
$*$-Hopf algebra
structures and bicovariant differential calculi, as was pointed out
in ref. \cite{Aschieri2} in the case of $GL_q(3) \rightarrow GL_q(2)$.
The reason why it works will be clarified below.
\sk
Let us explain what we mean by ``projection". To be specific, we
take again the case of $GL_q(3)$, generated by: i) the matrix elements
$\T{A}{B}$
\eq
\T{A}{B}=\left( \begin{array}{ccc}
\T{0}{0}&\T{0}{1}&\T{0}{2}\\
\T{1}{0}&\T{1}{1}&\T{1}{2}\\
\T{2}{0}&\T{2}{1}&\T{2}{2}\\
\end{array} \right )\equiv
\left( \begin{array}{ccc}
T_1&T_2&T_3\\T_4&T_5&T_6\\T_7&T_8&T_9\\
\end{array} \right ) \label{TGL}
\en
\noi ii) the identity $I$, and
the inverse $\Xi$ of
the $q$-determinant of $T$, defined by:
\eq \Xi~ \detq T=\detq T~\Xi=I\label{Xi} \en
\eq \detq T \equiv \sum_{\sigma} (-q)^{l(\sigma)}
\T{1}{\sigma(1)} \cdots
\T{n}{\sigma(n)} \label{qdet} \en
\noi where $l(\sigma)$ is the minimum number of
transpositions in the permutation
$\sigma$. Moreover, the matrix entries in (\ref{TGL})
satisfy the ``RTT" relations:
\eq
\R{AB}{EF} \T{E}{C} \T{F}{D} = \T{B}{F} \T{A}{E} \R{EF}{CD}
\label{RTTGL}
\en
\noi with $\R{AB}{CD}$ given by:
\eq
\R{AB}{CD}=\left(  \begin{array}{ccccccccc}
   q&0&0&0&0&0&0&0&0\\
   0&1&0&0&0&0&0&0&0\\
   0&0&1&0&0&0&0&0&0\\
   0&\lambda&0&1&0&0&0&0&0\\
0&0&0&0&q&0&0&0&0\\
0&0&0&0&0&1&0&0&0\\
0&0&\lambda&0&0&0&1&0&0\\
0&0&0&0&0&\lambda&0&1&0\\
0&0&0&0&0&0&0&0&q\\
\end{array} \right)
\en
\noi where $\lambda=q-q^{-1}~$. The reader can
verify that this $R$-matrix indeed has the form (\ref{RGL}).
We recall the co-structures of $GL_q(n+1)$,
i.e the coproduct
$\D$, the counit $\epsi$ and the coinverse $\kappa$:
\eqa
& & \D(\T{A}{B})=\T{A}{B} \otimes \T{B}{C}  \label{cos1} \\
& & \epsi (\T{A}{B})=\delta^A_B\\
& & \kappa(\T{A}{B})=\Ti{A}{B}\\
& &  \D (\detq T)=\detq T \otimes \detq T, ~~\D (\Xi)=\Xi \otimes \Xi,
{}~~\D(I)=I\otimes I\\
& &  \epsi (\detq T)=1,~~\epsi (\xi)=1,~~\epsi (I)=1\\
& &  \kappa (\detq T)=\xi,~~\kappa (\xi)=\detq T,~~\kappa (I)=I
\label{cos2}
\ena
\noi The quantum inverse of $\T{A}{B}$ is given by:
\eq
\Ti{A}{B}=\Xi~ (-q)^{A-B} t_B^{~A}
\en
\noi where $t_B^{~A}$ is the quantum minor, i.e. the quantum determinant
of the submatrix of $T$ obtained by removing the {\small $B$}-th row and
the {\small $A$}-th column.

A consistent $*$-structure of $GL_q(n)$ is given by \cite{FRT}:
\eq
(\T{A}{B})^*=\kappa(\T{B}{A}), \label{unitarity}
\en
\eq
\Xi^*=\detq T\label{starXi}
\en
\noi The unitarity condition (\ref{unitarity}) restricts $GL_q(n+1)$
to $U_q(n+1)$, while setting $\detq T =I$ restricts $GL_q(n+1)$ to
$SL_q(n+1)$. If both conditions hold we have $SU_q(n+1)$. For sake
of generality we consider $GL_q(n+1)$ rather than its restrictions,
but our discussion applies to $SL_q(n+1)$, $U_q(n+1)$ and $SU_q(n+1)$
as well. In the following we do not require (\ref{unitarity}) to
hold.
\sk
The projection of $GL_q(3)$ onto $GL_q(2)$ is defined by setting
\eq
T_2=T_3=T_4=T_7=0,~T_1=I \label{projT}
\en
\noi in (\ref{TGL}). The corresponding
left-invariant one-forms $\om$ and the $q$-Lie
algebra generators $\chi$ are set to ``zero" :
\eq
\om^2=\om^3=\om^4=\om^7=\om^1=0, \label{projom}
\en
\eq
\chi_2=\chi_3=\chi_4=\chi_7=\chi_1=0. \label{projchi}
\en
\noi (another equivalent projection would be given by $T_3=T_6=T_7=T_8=0,
{}~T_9=I$). Using (\ref{projom}) and (\ref{projchi}) in the differential
calculus of ref. \cite{Aschieri2}, one retrieves the bicovariant
differential calculus on $GL_q(2)$. The reason this projection works at
the quantum group level is of course the particular form of the
$R$-matrix in (\ref{RGL}), so that, for
example, the ``RTT" relations for
$GL_q(3)$ correctly reduce to those of $GL_q(2)$. Also, the $*$-Hopf
algebra structures project into those of $GL_q(2)$, as one can easily
deduce by substituting (\ref{projT}) into
(\ref{cos1})-(\ref{starXi}).
As we now discuss, the projection works also for the differential
calculi.
\sk
A bicovariant differential calculus \cite{Wor} on $GL_q(n+1)$
can be constructed in
terms of the corresponding $R$ matrix \cite{Jurco,Zumino,Aschieri}.
The basic object is the braiding matrix
\eq
\RRhat{A_1}{A_2}{D_1}{D_2}{C_1}{C_2}{B_1}{B_2}
\equiv  d^{F_2} d^{-1}_{C_2} \R{F_2B_1}{C_2G_1} \Rinv{C_1G_1}{E_1A_1}
    \Rinv{A_2E_1}{G_2D_1} \R{G_2D_2}{B_2F_2} \label{Lambda}
\en
which is used in the definition of the exterior product of
quantum left-invariant one forms $\ome{A}{B}$:
\eq
\ome{A_1}{A_2} \we \ome{D_1}{D_2}
\equiv \ome{A_1}{A_2} \otimes \ome{D_1}{D_2}
- \RRhat{A_1}{A_2}{D_1}{D_2}{C_1}{C_2}{B_1}{B_2}
\ome{C_1}{C_2} \otimes \ome{B_1}{B_2} \label{exteriorproduct}
\en
\noi and in the $q$-commutations of the quantum Lie
algebra generators $\cchi{A}{B}$:
\eq
\cchi{D_1}{D_2} \cchi{C_1}{C_2} - \RRhat{E_1}{E_2}{F_1}{F_2}
{D_1}{D_2}{C_1}{C_2} ~\cchi{E_1}{E_2} \cchi{F_1}{F_2} =
\CC{D_1}{D_2}{C_1}{C_2}{A_1}{A_2} \cchi{A_1}{A_2}
\label{qLie}
\en
\noi where the structure constants are explicitly given by:
\eq
\CC{A_1}{A_2}{B_1}{B_2}{C_1}{C_2} =\lam [- \de^{B_1}_{B_2}
\de^{A_1}_{C_1}
\de^{C_2}_{A_2} + \RRhat{B}{B}{C_1}{C_2}{A_1}{A_2}{B_1}{B_2}]. \label{CC}
\en
\noi and $\cchi{D_1}{D_2}
\cchi{C_1}{C_2} \equiv (\cchi{D_1}{D_2}  \otimes
\cchi{C_1}{C_2}) \D$, cf. ref.s \cite{Jurco,Zumino,Aschieri}.
\sk
The $d^A$ vector is defined by
\eq
\kappa^2(\T{A}{B})=\Dmat{A}{C} \T{C}{D} \Dmatinv{D}{B}=d^A d^{-1}_B
\T{A}{B} \label{defD}
\en
\noi For $GL_q(n+1)$ we have
$d^A=q^{2A-1}$ (cf. \cite{FRT}). In the case
of $GL_q(3)$, $d^0=q$, $d^1=q^3$, $d^2=q^5$.
\sk
The braiding matrix $\La$  and
the structure constants $\Cb$ defined in (\ref{Lambda}) and
(\ref{CC}) satisfy
the conditions

\eqa
& & \C{ri}{n} \C{nj}{s}-\L{kl}{ij} \C{rk}{n} \C{nl}{s} =
\C{ij}{k} \C{rk}{s}
{}~~\mbox{({\sl q}-Jacobi identities)} \label{bic1}\\
& & \L{nm}{ij} \L{ik}{rp} \L{js}{kq}=\L{nk}{ri} \L{ms}{kj}
\L{ij}{pq}~~~~~~~~~\mbox{(Yang--Baxter)} \label{bic2}\\
& & \C{mn}{i} \L{ml}{rj} \L{ns}{lk} + \L{il}{rj} \C{lk}{s} =
\L{pq}{jk} \L{il}{rq} \C{lp}{s} + \C{jk}{m} \L{is}{rm}
\label{bic3}\\
& & \C{rk}{m} \L{ns}{ml} = \L{ij}{kl} \L{nm}{ri} \C{mj}{s}
\label{bic4}
\ena

\noi where the index pairs ${}_A^{~B}$ and ${}^A_{~B}$
have been replaced by the indices ${}^i$ and ${}_i$ respectively.
These are the so-called ``bicovariance conditions", see ref.s
\cite{Wor,Bernard,Aschieri},
necessary in order to have a consistent bicovariant differential
calculus.
\sk
By using (\ref{RGL}) in (\ref{Lambda}) and (\ref{CC}), one finds that
$\RRhat{A_1}{A_2}{D_1}{D_2}{b_1}{b_2}{c_1}{c_2}=0$ unless {\small
$A_1=a_1, A_2=a_2, D_1=d_1, D_2=d_2$}, and
$\CC{c_1}{c_2}{b_1}{b_2}{D_1}{D_2}=0$
unless {\small $D_1=d_1,D_2=d_2$}. As a consequence
$\RRhat{a_1}{a_2}{d_1}{d_2}
{b_1}{b_2}{c_1}{c_2}$ and $\CC{c_1}{c_2}{b_1}{b_2}{d_1}{d_2}$
satisfy {\sl by themselves} the bicovariance conditions
(\ref{bic1})-(\ref{bic4}). This explains why the projection
from $GL_q(n+1)$ to $GL_q(n)$:
\eq
\T{0}{a}=\T{a}{0}=0,~\T{0}{0}=I,
\en
\eq
\ome{0}{a}=\ome{a}{0}=\ome{0}{0}=0,
\en
\eq
\cchi{0}{a}=\cchi{a}{0}=\cchi{0}{0}=0
\en
\noi leads to a consistent bicovariant calculus for $GL_q(n)$.
\sk
So far we have seen how to obtain $GL_q(n)$, together with its
$*$-Hopf algebra structure and bicovariant differential calculus
from the ``mother " structures of $GL_q(n+1)$. This is not so exciting,
but suggests a way to obtain inhomogeneous quantum groups via
another kind of projection.
\sk
Indeed, consider
\eq
\T{0}{a}=0,~\T{a}{0}=x^a,~\T{0}{0}=u \label{projTi}
\en
\noi (note that $\T{0}{0}$ is not set to the identity any more),
together with
\eq
\ome{0}{a}=\ome{0}{0}=0 \label{projomi}
\en
\eq
\cchi{0}{a}=\cchi{0}{0}=0 \label{projchii}
\en
\noi
The projection (\ref{projTi}) yields the quantum group $IGL_q(n)$,
generated by $\T{a}{b}$, $x^a$, $u$, $v$ (the inverse of $u$, i.e.
$uv=vu=I$),
$\xi$ (the inverse of $\detq \T{a}{b}$) and the identity $I$. The
commutation relations of these elements can be read off the ``RTT"
relations (\ref{RTTGL}) for $GL_q(n+1)$ after using (\ref{RGL}) and
(\ref{projTi}):
\eq
\R{ab}{ef} \T{e}{c} \T{f}{d} = \T{b}{f} \T{a}{e} \R{ef}{cd}
\label{RTTGLn}
\en
\eq
x^a \T{b}{c}=\R{ba}{ef} \T{e}{c} x^f \label{xT}
\en
\eq
\A{ab}{cd} x^c x^d=0 \label{Axx}
\en
\eq
\T{a}{b} u= u \T{a}{b} \label{Tu}
\en
\eq
\T{a}{b} v= v \T{a}{b} \label{Tv}
\en
\eq
x^a u=\qm u x^a \label{xu}
\en
\eq
x^a v=q v x^a \label{xv}
\en
\noi the $A$ matrix being the $q$-generalization of the antisymmetrizer:
\eq
A={{qI-\Rh}\over{q+\qm}} \label{defA}
\en
\noi where $\Rhat{ab}{cd} \equiv \R{ba}{cd}$.

The ``projected"
quantum determinant $\detq \T{A}{B} = u \det_q \T{a}{b}$
and its inverse $\Xi=v\xi$ are central.
\sk
The
Hopf algebra co-structures, consistent with the commutation rules,
are deduced from those of $GL_q(n+1)$ by
simply substituting (\ref{projTi}) into (\ref{cos1})-(\ref{unitarity}):
\eqa
& &\D(\T{a}{b})=\T{a}{c}
\otimes \T{c}{b},~~\D (I)=I\otimes I,\label{DT}\\
& & \D(x^a)=\T{a}{b} \otimes x^b + x^a \otimes u \label{Dx}\\
& & \D(u)=u\otimes u,~~\D(v)=v\otimes v \label{Du}\\
& & \D(\detq T)=\detq T \otimes \detq T,~~\D(\xi)=\xi \otimes \xi
\label{Ddetq}
\ena
\eqa
& & \epsi(\T{a}{b})=\de^a_b,~~\epsi (I)=1,\label{epsiT}\\
& & \epsi(x^a)=0 \label{epsix}\\
& & \epsi(u)=\epsi(v)=1 \label{epsiu}\\
& & \epsi(\detq T)=\epsi(\xi)=1 \label{epsidetq}
\ena
\eqa
& & \kappa(\T{a}{b})=\Ti{a}{b},~~\kappa(I)=I, \label{kappaT}\\
& & \kappa(x^a)=-\kappa(\T{a}{b}) x^b v\label{kappax}\\
& & \kappa(u)=v,~~\kappa(v)=u  \label{kappau}\\
& & \kappa(\detq T)=\xi,~~\kappa(\xi)=\detq T \label{kappadet}
\ena
After using (\ref{projomi}) and (\ref{projchii}) do we obtain
a consistent bicovariant differential calculus for
the quantum group $IGL_q(n)$ ? The answer is yes. Indeed
consider the $q$-Lie algebra (\ref{qLie}) of $GL_q(n+1)$. Using
the decomposition (\ref{RGL}) for $\R{AB}{CD}$ we find
\eqa
 \cchi{c_1}{c_2}\cchi{b_1}{b_2}&-&
\LL{a_1}{a_2}{d_1}{d_2}{c_1}{c_2}{b_1}{b_2}~
\cchi{a_1}{a_2} \cchi{d_1}{d_2}=\linv [-\de^{b_1}_{b_2}
\de^{c_1}_{d_1} \de^{d_2}_{c_2} + \LL{a}{a}{d_1}{d_2}{c_1}{c_2}{b_1}
{b_2}] \cchi{d_1}{d_2}\label{qLie1}\\
\cchi{c_1}{0} \cchi{b_1}{b_2}&-&\Rinv{c_1b_1}{e_1a_1}
\Rinv{a_2e_1}{b_2d_1} ~\cchi{a_1}{a_2} \cchi{d_1}{0}=\nonumber\\
& &\linv [-\de^{b_1}_{b_2} \de^{c_1}_{d_1} + \Rinv{c_1b_1}{e_1a}
\Rinv{ae_1}{b_2d_1} ] \cchi{d_1}{0} \label{qLie2}\\
\cchi{c_1}{c_2} \cchi{b_1}{0} &-& (q-\qm) d^{g_2} d^{-1}_{c_2}
\R{g_2b_1}{c_2g_1} \Rinv{c_1g_1}{e_1a_1} \Rinv{a_2e_1}
{g_2d_1} ~\cchi{a_1}{a_2} \cchi{d_1}{\bullet}- \nonumber\\
& &d^{d_2} d^{-1}_{c_2}
\R{d_2b_1}{c_2g_1} \Rinv{c_1g_1}{d_1a_1} \cchi{a_1}{0}
\cchi{d_1}{d_2}=\nonumber\\
&=&d^{g_2} d^{-1}_{c_2} \R{g_2b_1}{c_2g_1}
\Rinv{c_1g_1}{e_1a}\Rinv{ae_1}{g_2d_1} \cchi{d_1}{0}
\label{qLie3}\\
\cchi{c_1}{0}\cchi{b_1}{0}&-&q~\Rinv{c_1b_1}{d_1a_1}
{}~\cchi{a_1}{0} \cchi{d_1}{0}=0 \label{qLie4}
\ena
\noi where $\LL{a_1}{a_2}{d_1}{d_2}{c_1}{c_2}{b_1}{b_2}$
is the
braiding matrix
of $GL_q(n)$, given in (\ref{L1}), so that the commutations in
(\ref{qLie1}) are those of the $q$-subalgebra $GL_q(n)$.
Note that the $\qone$ limit on the right hand sides of (\ref{qLie1})
and (\ref{qLie2}) is finite, since the terms in square parentheses
are a (finite) series in $q-\qm$, and the $0-th$ order part vanishes
(see \cite{Aschieri}, eq. (5.55)). We have
written here only a subset $X$ of the commutation
relations (\ref{qLie}). This subset involves only the
$\cchi{a}{b}$ and $\cchi{a}{0}$ generators, and
closes on itself. The $\La$ and $\Cb$ components entering
(\ref{qLie1})-(\ref{qLie4}) are
\eqa
& &\LL{a_1}{a_2}{d_1}{d_2}{c_1}{c_2}{b_1}{b_2}=
d^{f_2} d^{-1}_{c_2} \R{f_2b_1}{c_2g_1} \Rinv{c_1g_1}{e_1a_1}
    \Rinv{a_2e_1}{g_2d_1} \R{g_2d_2}{b_2f_2} \label{L1}\\
& &\LL{a_1}{0}{d_1}{d_2}{c_1}{c_2}{b_1}{0}=
d^{d_2} d^{-1}_{c_2} \R{d_2b_1}{c_2g_1} \Rinv{c_1g_1}{d_1a_1}
     \label{L6}\\
& &\LL{a_1}{a_2}{d_1}{0}{c_1}{0}{b_1}{b_2}=
 \Rinv{c_1b_1}{e_1a_1} \Rinv{a_2e_1}{b_2d_1}  \label{L7}\\
& &\LL{a_1}{a_2}{d_1}{0}{c_1}{c_2}{b_1}{0}=
(q-\qm) d^{g_2} d^{-1}_{c_2} \R{g_2b_1}{c_2g_1} \Rinv{c_1g_1}{e_1a_1}
    \Rinv{a_2e_1}{g_2d_1} \label{L8}\\
& &\LL{a_1}{0}{d_1}{0}{c_1}{0}{b_1}{0}=
   q \Rinv{c_1b_1}{d_1a_1}     \label{L9}
\ena
\eqa
& &\CC{c_1}{c_2}{b_1}{b_2}{d_1}{d_2}=
{\rm structure~constants~of~}GL_q(n)\\
& &\CC{c_1}{0}{b_1}{b_2}{d_1}{0}=
\linv [-\de^{b_1}_{b_2} \de^{c_1}_{d_1} + \Rinv{c_1b_1}{e_1a}
\Rinv{ae_1}{b_2d_1} ] \\
& &\CC{c_1}{c_2}{b_1}{0}{d_1}{0}=
d^{g_2} d^{-1}_{c_2} \R{g_2b_1}{c_2g_1}
\Rinv{c_1g_1}{e_1a}\Rinv{ae_1}{g_2d_1} \label{C3}
\ena
\noi This result was already found in \cite{Cas1} without using
the projection discussed in the present Letter. We repeat now
the same reasoning as in \cite{Cas1}: the
components given in (\ref{L1})-(\ref{C3}) are the only non-vanishing
$\LL{A_1}{A_2}{D_1}{D_2}{C_1}{C_2}{B_1}{B_2}$ components
and the only non-vanishing
$\CC{C_1}{C_2}{B_1}{B_2}{D_1}{D_2}$ components with indices
{\small $C_1,C_2,B_1,B_2$} corresponding to
the subset $X$. Because of this, they satisfy {\sl by themselves}
the bicovariance conditions, as
the sums in (\ref{bic1})-(\ref{bic4}) do not
involve other components. Then
(\ref{qLie1})-(\ref{qLie4}) defines a bicovariant
quantum Lie algebra, and a consistent differential calculus can be set
up, based on a $\La$ tensor whose only nonvanishing components are
(\ref{L1})-(\ref{L9}).
\sk
Finally, we consider the complexification $IGL_q(n,\Cb)$.
This we obtain as the projection of the complex $q$-group
$GL_q(n+1,\Cb)$. Let us recall how to construct $GL_q(n+1,\Cb)$ from
$GL_q(n+1)$ \cite{Drabant}. Using the $*$-structure on
$GL_q(n+1)$, one introduces the
conjugated elements
\eq
\Th{A}{B} \equiv [\kappa(\T{B}{A})]^*
\en
The complex conjugate version of (\ref{RTTGL})
yields the $R {\hat T} {\hat T}$
relation:
\eq
\R{AB}{EF} \Th{E}{C} \Th{F}{D} =
\Th{B}{F} \Th{A}{E} \R{EF}{CD} \label{RThTh}
\en
\noi whereas the commutations between $\T{A}{B}$ and $\Th{A}{B}$ can be
defined to be
\eq
\R{AB}{EF} \Th{E}{C} \T{F}{D} =
\T{B}{F} \Th{A}{E} \R{EF}{CD} \label{RThT}
\en

An $RTT$-formulation
for the complexified quantum group $GL_q(n+1,\Cb)$ can be found by
defining the matrix
$\T{J}{K}$:
\eq
\T{J}{K}=\left( \begin{array}{cc}
\T{A}{B}&0\\
0&\T{\Ab}{\Bb}\\
\end{array} \right )
\en
\noi where
\eq
\T{\Ab}{\Bb} \equiv \Th{A}{B}
\en
\noi with the index
convention {\small $J=(A,\Ab)$}. Then the $\Rbo T T$ relation
\eq
\Rb{IJ}{MN} \T{M}{K} \T{N}{L}=\T{J}{N} \T{I}{M} \Rb{MN}{KL}
\label{RTTcomplex}
\en
\noi with
\eq
\Rb{IJ}{KL}=\left( \begin{array}{cccc}
R&0&0&0\\
0&(R^+)^{-1}&0&0\\
0&0&R&0\\
0&0&0&R \\
\end{array} \right )\label{Rcomplex}
\en
\noi (indices {\small $IJ$} ordered as {\small $AB,A{\bar B},
{\bar A}B,{\bar A}{\bar B}$}, and $\Rp{AB}{CD} \equiv \R{BA}{DC}$)
reproduces
the commutations (\ref{RTTGL}), (\ref{RThTh}) and (\ref{RThT}).
A bicovariant calculus on $GL_q(n+1,\Cb)$
can be set up in terms of the $\Rbo$ matrix,
via the standard formula given before for the braiding matrix $\Lambda$.
\sk
We define now the projection of $GL_q(n+1,\Cb)$ onto the complexified
inhomogeneous quantum group $IGL_q(n,\Cb)$ by taking the complexified
version of (\ref{projTi}):
\eq
\T{0}{a}=0,~\T{a}{0}\equiv x^a,~\T{0}{0}\equiv u,~
\T{\bar 0}{\bar a}=0,~\T{\bar a}{\bar 0}\equiv x^{\bar a},
{}~\T{\bar 0}{\bar 0} \equiv \ub \label{projTic}
\en
\noi Then $IGL_q(n+1,\Cb)$ is defined
as the algebra $\cal A$ freely
generated by the elements $\T{a}{b}$, $\T{\ab}{\bb}$, $x^a$, $x^{\ab}$,
$u$ and its inverse $v$, $\ub$ and its inverse $\vb$, and the
inverses $\xi$, $\xib$ of the $q$-determinants
$\detq \T{a}{b}$, $\detq \T{\ab}{\bb}$. The ``projected" $q$-determinants
$u \detq \T{a}{b}$, $\ub \detq \T{\ab}{\bb}$ and their inverses $v\xi$,
$\vb \xib$ are
central. The commutations of these elements are deduced from
(\ref{RTTGL}), (\ref{RThTh}) and (\ref{RThT}) after use of
(\ref{projTic}); they are given therefore by (\ref{RTTGLn})-(\ref{xv})
for $TT$ and ${\hat T}{\hat T}$ commutations, whereas the ${\hat T} T$
commutations
are:
\eq
\R{ab}{ef} \T{\bar e}{\bar c} \T{f}{d} = \T{b}{f} \T{\bar a}{\bar e}
\R{ef}{cd}
\label{RTbarT}
\en
\eq
\R{ab}{ef} \T{\bar e}{\bar c} x^f=x^b \T{\bar a}{\bar c}
\en
\eq
\T{\bar a}{\bar c} u=u \T{\bar a}{\bar c},~~
\T{\bar a}{\bar c} v=v \T{\bar a}{\bar c}
\en
\eq
\R{ab}{ef} x^{\bar e} \T{f}{d} = (q-\qm) x^b \T{\bar a}{\bar d} +
\T{b}{d} x^{\bar a}
\en
\eq
\R{ab}{ef} x^{\bar e} x^f=q x^b x^{\bar a}
\en
\eq
x^{\bar a} u + (q-\qm) \ub x^a=q u x^{\bar a}
\en
\eq
v x^{\bar a} + (q-\qm)v \ub x^a v=q x^{\bar a} v
\en
\eq
\ub \T{a}{b}=\T{a}{b} \ub,~~\vb \T{a}{b} = \T{a}{b} \vb
\en
\eq
\ub x^a=q x^a \ub,~~\vb x^a = \qm x^a \vb
\en
\eq
\ub u=u \ub,~~\ub v=v \ub,~~\vb u = u\vb,~~\vb v= v \vb
\en
The co-structures of the conjugated elements $\T{\Ab}{\Bb}$ are given
by the same formulas (with barred indices) as in
(\ref{cos1})-(\ref{cos2}).
\sk
The bicovariant differential calculus on $GL_q(n+1,\Cb)$ is found
by the usual procedure, described in formulas (\ref{Lambda})-(\ref{CC}),
after replacing {\small $A,B...$} indices by {\small $I,J...$} indices.
Here again we find a subset $X$ of the $q$-Lie algebra of $GL_q(n+1,\Cb)$
that closes on itself, and allows therefore a consistent projection
onto a bicovariant differential calculus for $IGL_q(n,\Cb)$. This
subset is given by (\ref{qLie1})-(\ref{qLie4}) and:
\eqa
 \cchi{\bar c_1}{\bar c_2}\cchi{\bar b_1}{\bar b_2}&-&
\LL{a_1}{a_2}{d_1}{d_2}{c_1}{c_2}
{b_1}{b_2}~
\cchi{\bar a_1}{\bar a_2} \cchi{\bar d_1}{\bar d_2}=\linv
[-\de^{b_1}_{b_2}
\de^{c_1}_{d_1} \de^{d_2}_{c_2} + \LL{a}{a}{d_1}
{d_2}{c_1}{c_2}{b_1}
{b_2}] \cchi{\bar d_1}{\bar d_2}\label{qLie1c}\\
\cchi{\bar 0}{\bar c_2} \cchi{\bar b_1}{\bar b_2}&-& d^{f_2}
d^{-1}_{c_2} \R{f_2b_1}{c_2a_1}
\R{a_2d_2}{b_2f_2} ~\cchi{\bar a_1}{\bar a_2} \cchi{\bar 0}{\bar d_2}=
\nonumber\\
& &\linv [-\de^{b_1}_{b_2} \de^{d_2}_{c_2} + d^{f_2} d^{-1}_{c_2}
\R{f_2b_1}{c_2a}
\R{ad_2}{b_2f_2} ] \cchi{\bar 0}{\bar d_2} \label{qLie2c}\\
\cchi{\bar c_1}{\bar c_2} \cchi{\bar 0}{\bar b_2} &+
&(q-\qm) \de^{c_1}_{a_1}\R{a_2d_2}{b_2c_2} ~\cchi{\bar a_1}{\bar a_2}
\cchi{\bar 0}{\bar d_2}-
 \nonumber\\
&-&\Rinv{a_2c_1}{g_2d_1} \R{g_2d_2}{b_2f_2} \cchi{\bar 0}{\bar a_2}
\cchi{\bar d_1}{\bar d_2}= \R{c_1d_2}{b_2c_2} \cchi{\bar 0}{\bar d_2}
\label{qLie3c}\\
\cchi{\bar 0}{\bar c_2}\cchi{\bar 0}{\bar b_2}&-&\qm~\R{a_2d_2}{b_2c_2}
{}~\cchi{\bar 0}{\bar a_2} \cchi{\bar 0}{\bar d_2}=0 \label{qLie4c}
\ena
\eqa
 \cchi{c_1}{c_2}\cchi{\bar b_1}{\bar b_2}&+&
(q-\qm) d^0 d^{-1}_{c_2} \de^{b_1}{c_2} \Rinv{a_2c_1}{b_2d_1}
{}~\cchi{\bar 0}{\bar a_2} \cchi{d_1}{0}-\nonumber\\
&-& d^{f_2} d^{-1}_{c_2} \Rinv{b_1f_2}{g_1c_2} \R{g_1c_1}{a_1e_1}
    \Rinv{a_2e_1}{g_2d_1} \R{g_2d_2}{b_2f_2}~
\cchi{\bar a_1}{\bar a_2} \cchi{d_1}{d_2}=0 \label{qLie1mixed}\\
 \cchi{\bar c_1}{\bar c_2}\cchi{b_1}{b_2}&-&
(q-\qm) d^{f_2} d^{-1}_{c_2} \R{f_2b_1}{c_2g_1} \Rinv{c_1g_1}{g_2a_1}
\Rinv{d_2g_2}{f_2b_2}
\cchi{a_1}{0} \cchi{\bar 0}{\bar d_2}-\nonumber\\
&-& d^{f_2} d^{-1}_{c_2} \R{f_2b_1}{c_2g_1} \Rinv{c_1g_1}{e_1a_1}
    \R{e_1a_2}{d_1g_2} \Rinv{d_2g_2}{f_2b_2}~
\cchi{a_1}{a_2} \cchi{\bar d_1}{\bar d_2}=0 \label{qLie2mixed}\\
\cchi{c_1}{c_2} \cchi{\bar 0}{\bar b_2} &-&
\Rinv{a_2c_1}{g_2d_1} \R{g_2d_2}{b_2c_2} ~\cchi{\bar 0}{\bar a_2}
\cchi{d_1}{d_2}=0 \label{qLie3mixed}\\
\cchi{\bar 0}{\bar c_2} \cchi{b_1}{b_2} &-& d^{f_2} d^{-1}_{c_2}
\R{f_2b_1}{c_2a_1} \Rinv{d_2a_2}{f_2b_2} ~\cchi{a_1}{a_2} \cchi{\bar 0}
{\bar d_2}=0 \label{qLie4mixed}\\
\cchi{c_1}{0} \cchi{\bar b_1}{\bar b_2}&-&\R{b_1c_1}{a_1e_1}
\Rinv{a_2e_1}{b_2d_1} ~\cchi{\bar a_1}{\bar a_2} \cchi{d_1}{0}=0
\label{qLie5mixed}\\
\cchi{\bar c_1}{\bar c_2} \cchi{b_1}{0} &-& d^{d_2} d^{-1}_{c_2}
\R{d_2b_1}{c_2g_1}\Rinv{c_1g_1}{d_1a_1}~ \cchi{a_1}{0}
\cchi{\bar d_1}{\bar d_2} = 0 \label{qLie6mixed}\\
\cchi{c_1}{0}\cchi{\bar 0}{\bar b_2}&-&q^{-1}~\Rinv{a_2c_1}{b_2d_1}
{}~\cchi{\bar 0}{\bar a_2} \cchi{d_1}{0}=0 \label{qLie7mixed}\\
\cchi{\bar 0}{\bar c_2} \cchi{b_1}{0}&-&qd^{d_2} d^{-1}_{c_2}
\R{d_2b_1}{c_2a_1} ~\cchi{a_1}{0} \cchi{\bar 0}{\bar d_2}=0
\label{qLie8mixed}
\ena

The commutations in
(\ref{qLie1})-(\ref{qLie4}) and (\ref{qLie1c})-(\ref{qLie8mixed})
are those of the two $q$-commuting
subalgebras $GL_q(n)$.
Again we call $X$ this subset of the $q$-Lie algebra commutation
relations of $GL_q(n+1,\Cb)$. It closes on the
generators $\cchi{a}{b}$, $\cchi{a}{0}$, $\cchi{\bar a}{\bar b}$,
$\cchi{\bar 0}{\bar b}$. The $\La$ and $\Cb$ components entering
the subset $X$ are
given by (\ref{L1})-(\ref{C3}) and by
\eqa
& &\LL{\bar a_1}{\bar a_2}{\bar d_1}{\bar d_2}{\bar c_1}{\bar c_2}
{\bar b_1}{\bar b_2}=\LL{a_1}{a_2}{d_1}{d_2}{c_1}{c_2}{b_1}{b_2}
\label{L1c}\\
& &\LL{\bar 0}{\bar a_2}{\bar d_1}{\bar d_2}{\bar c_1}{\bar c_2}
{\bar 0}{\bar b_2}=
\Rinv{a_2c_1}{g_2d_1} \R{g_2d_2}{b_2f_2}
 \label{L2c}\\
& &\LL{\bar a_1}{\bar a_2}{\bar 0}{\bar d_2}{\bar c_1}{\bar c_2}
{\bar 0}{\bar b_2}=-(q-\qm) \de^{c_1}_{a_1} \R{a_2d_2}{b_2c_2}
\label{L3c}\\
& &\LL{\bar a_1}{\bar a_2}{\bar 0}{\bar d_2}{\bar 0}{\bar c_2}{\bar b_1}
{\bar b_2}=
 d^{f_2} d^{-1}_{c_2} \R{f_2b_1}{c_2a_1} \R{a_2d_2}{b_2f_2}
\label{L4c}\\
& &\LL{\bar 0}{\bar a_2}{\bar 0}{\bar d_2}{\bar 0}{\bar c_2}
{\bar 0}{\bar b_2}=
\qm \R{a_2d_2}{b_2c_2}     \label{L5c}
\ena

\eqa
& &\LL{\bar a_1}{\bar a_2}{d_1}{d_2}{c_1}{c_2}{\bar b_1}{\bar b_2}=
d^{f_2} d^{-1}_{c_2} \Rinv{b_1f_2}{g_1c_2} \R{g_1c_1}{a_1e_1}
    \Rinv{a_2e_1}{g_2d_1} \R{g_2d_2}{b_2f_2} \label{L1mixed}\\
& &\LL{\bar 0}{\bar a_2}{d_1}{0}{c_1}{c_2}{\bar b_1}{\bar b_2}=
-d^0 d^{-1}_{c_2} (q-\qm) \de^{b_1}_{c_2}\Rinv{a_2c_1}{b_2d_1}
\label{L2mixed}\\
& &\LL{a_1}{a_2}{\bar d_1}{\bar d_2}{\bar c_1}{\bar c_2}{b_1}{b_2}=
d^{f_2} d^{-1}_{c_2} \R{f_2b_1}{c_2g_1} \Rinv{c_1g_1}{e_1a_1}
    \R{e_1a_2}{d_1g_2} \Rinv{d_2g_2}{f_2b_2} \label{L3mixed}\\
& &\LL{a_1}{0}{\bar 0}{\bar d_2}{\bar c_1}{\bar c_2}{b_1}{b_2}=
(q-\qm) d^{f_2} d^{-1}_{c_2}
\R{f_2b_1}{c_2g_1} \Rinv{c_1g_1}{g_2a_1} \Rinv{d_2g_2}{f_2b_2}
\label{L4mixed}\\
& &\LL{\bar 0}{\bar a_2}{d_1}{d_2}{c_1}{c_2}{\bar 0}{\bar b_2}=
\Rinv{a_2c_1}{g_2d_1} \R{g_2d_2}{b_2c_2}
     \label{L5mixed}\\
& &\LL{a_1}{a_2}{\bar 0}{\bar d_2}{\bar 0}{\bar c_2}{b_1}{b_2}=
 d^{f_2} d^{-1}_{c_2} \R{f_2b_1}{c_2a_1} \Rinv{d_2a_2}{f_2b_2}
\label{L6mixed}\\
& &\LL{\bar a_1}{\bar a_2}{d_1}{0}{c_1}{0}{\bar b_1}{\bar b_2}=
d^{f_2} d^{-1}_{c_2} \R{b_1c_1}{a_1e_1} \Rinv{a_2e_1}{b_2d_1}
\label{L7mixed}\\
& &\LL{a_1}{0}{\bar d_1}{\bar d_2}{\bar c_1}{\bar c_2}{b_1}{0}=
  d^{d_2} d^{-1}_{c_2} \R{d_2b_1}{c_2g_1} \Rinv{c_1g_1}{d_1a_1}
\label{L8mixed}\\
& &\LL{\bar 0}{\bar a_2}{d_1}{0}{c_1}{0}{\bar 0}{\bar b_2}=
   \qm \Rinv{a_2c_1}{b_2d_1}     \label{L9mixed}\\
& &\LL{a_1}{0}{\bar 0}{\bar d_2}{\bar 0}{\bar c_2}{b_1}{0}=
qd^{d_2} d^{-1}_{c_2} \R{d_2b_1}{c_2a_1} \label{L10mixed}
\ena
\eqa
& &\CC{\bar c_1}{\bar c_2}{\bar b_1}{\bar b_2}{\bar d_1}{\bar d_2}=
\CC{c_1}{c_2}{b_1}{b_2}{d_1}{d_2}\\
& &\CC{\bar 0}{\bar c_2}{\bar b_1}{\bar b_2}{\bar 0}{\bar d_2}=
\linv [-\de^{b_1}_{b_2} \de^{d_2}_{c_2} + d^{f_2} d^{-1}_{c_2}
\R{f_2b_1}{c_2a}
\R{ad_2}{b_2f_2} ] \\
& &\CC{\bar c_1}{\bar c_2}{\bar 0}{\bar b_2}{\bar 0}{\bar d_2}=
-\R{c_1d_2}{b_2c_2}
\ena

\noi Again we find that these are the only non-vanishing
$\LL{A_1}{A_2}{D_1}{D_2}{C_1}{C_2}{B_1}{B_2}$ components
and the only non-vanishing
$\CC{C_1}{C_2}{B_1}{B_2}{D_1}{D_2}$ components with indices
{\small $C_1,C_2,B_1,B_2$} corresponding to
the subset $X$. By the same reasoning used in the case of $IGL_q(n)$,
we conclude that
(\ref{qLie1})-(\ref{qLie4}),(\ref{qLie1c})-(\ref{qLie8mixed}) define
a bicovariant
quantum Lie algebra, and a consistent differential calculus can be set
up, based on a $\La$ tensor whose only nonvanishing components are
(\ref{L1})-(\ref{L9}), (\ref{L1c})-(\ref{L10mixed}). This differential
calculus is obtained from the one of $GL_q(n+1,\Cb)$ by setting:
\eq
\ome{0}{a}=\ome{0}{0}=\ome{\bar 0}{\bar a}=\ome{\bar 0}{\bar 0}=0
\label{projomic}
\en
\eq
\cchi{0}{a}=\cchi{0}{0}=\cchi{\bar 0}{\bar a}=\cchi{\bar 0}{\bar 0}=0
\label{projchiic}
\en
\sk
In the Table we present the $q$-Lie algebra for $IGL_q(2,\Cb)$,
i.e. the quantum Poincar\'e group with the addition of two dilatations,
generated by $\cchi{1}{1}+\cchi{2}{2}$ and $\cchi{\bar 1}{\bar 1}+
\cchi{\bar 2}{\bar 2}$. The generators $\cchi{a}{b}$
and $\cchi{\ab}{\bb}$ close on the $q$-Lie algebra of
$GL_q(2,\Cb)$, while $\cchi{a}{0}$ and $\cchi{\bar 0}{\bar b}$
are the four $q$-momentum generators. To obtain $ISL_q(2,\Cb)$ we
must require $ u \detq \T{a}{b}=
\ub \detq \T{\ab}{\bb}=I$. This implies the relation
\eq
\cchi{1}{1}+\cchi{2}{2}-(q-\qm)\cchi{1}{1} \cchi{2}{2} + q^2
(q-\qm) \cchi{1}{2}\cchi{2}{1}=0 \label{chirelation}
\en
\noi (cf. ref. \cite{Zumino}) and a similar one for barred generators,
which reduce the number of independent generators from 12 to 10.
\sk
The co-structures of $\cchi{I}{J}$ are given by:
\eqa
& &\Dp(\cchi{I}{J})=\Ip \otimes \cchi{I}{J} + \cchi{K}{L} \otimes
\ff{K}{LI}{J}\\
& &\ep (\cchi{I}{J})=0\\
& &\kp (\cchi{I}{J})=-\cchi{K}{L}~ \ff{K}{LI}{J}
\ena
\noi where
\eq
\ff{K}{LI}{J} \equiv \kp(\Lp{I}{K}) \Lm{L}{J}
\en
\noi and the functionals $\Lpm{I}{J}$ are defined below. A
detailed account of the bicovariant differential calculus
on the quantum Poincar\'e group is given in ref. \cite{Cas2}.
\sk
{\sl Note 1:} we have chosen $\Rb{\Ab\Bb}{{\bar C}\Db}=\R{AB}{CD}$ in
 (\ref{Rcomplex}). In fact, another choice is possible, i.e.
$\Rb{\Ab\Bb}{{\bar C}\Db}=\Rpinv{AB}{CD}$, since it reproduces the same
commutations (\ref{RThTh}). This last choice is favoured in ref.
\cite{Drabant}. However a consistent projection on $IGL_q(n,\Cb)$
does not seem to exist in this case. Note that our choice
(\ref{Rcomplex}) is still consistent with a $*$-structure on the space
of regular functionals. Indeed a conjugation on the functionals
$\Lpm{A}{B}$ can be defined as:
\eq
[\Lpm{A}{B}]^{\dagger} (a) \equiv \overline{[\Lmp{B}{A} (a^*)]}
\en
\noi where $a \in GL_q(n+1,\Cb)$ and the bar indicates the usual
conjugation on $\Cb$. We recall that these functionals
are defined by their action on the group elements:
\eq
\Lpm{I}{J} (\T{K}{L})=\Rbpm{IK}{JL}
\en
\noi Then
\eq
\Lpm{\Ab}{\Bb}=[\Lpm{A}{B}]^{\dagger}
\en
\sk
{\sl Note 2:} the right-hand sides of eq.s (\ref{qLie1mixed}) and
(\ref{qLie2mixed}) vanish because of the identities:
\eq
d^{f_2} d^{-1}_{c_2} \Rinv{b_1f_2}{gc_2} \R{gd_2}{b_2f_2} =
\de^{d_2}_{c_2} \de^{b_1}_{b_2}
\en
\eq
d^{f_2} d^{-1}_{c_2} \Rinv{f_2b_1}{c_2g} \R{d_2g}{f_2b_2} =
\de^{d_2}_{c_2} \de^{b_1}_{b_2}
\en
\noi valid for any $GL_q(n)$ $R$-matrix.
\sk
{\sl Note 3:} the quantum Lorentz group $SL_q(2,\Cb)$ is obviously
contained in the $q$-Poincar\'e group $ISL_q(2,\Cb)$. This inclusion
holds also for the corresponding $q$-Lie algebras, since $\cchi{a}{b}$
and $\cchi{\ab}{\bb}$ close on the quantum Lorentz $q$-Lie algebra, cf.
the Table. This fact is of relevance for the construction of a
$q$-Minkowski spacetime as the quantum coset space
$q$-Poincar\'e / $q$-Lorentz.

\vfill\eject

\centerline{\bf Table}
\centerline{\sl The bicovariant $q$-Lie algebra of the quantum
Poincar\'e group}
\sk
\[
\begin{array}{l}
\cchi{1}{1} \cchi{1}{2}-\cchi{1}{2} \cchi{1}{1}+(1-q^2)
\cchi{1}{2} \cchi{2}{2}=q\cchi{1}{2}\\
\cchi{1}{1} \cchi{2}{1}-\cchi{2}{1} \cchi{1}{1}-(1-q^2)
\cchi{2}{2} \cchi{2}{1}=-q\cchi{2}{1}\\
\cchi{1}{1} \cchi{2}{2}-\cchi{2}{2} \cchi{1}{1}=0\\
\cchi{1}{2} \cchi{2}{1}-\cchi{2}{1} \cchi{1}{2}+(1-q^2)
\cchi{2}{2} \cchi{1}{1}-(1-q^2) \cchi{2}{2} \cchi{2}{2}=
q(\cchi{1}{1} - \cchi{2}{2})\\
\cchi{1}{2} \cchi{2}{2}- q^2 \cchi{2}{2} \cchi{1}{2}=q\cchi{1}{2}\\
\cchi{2}{1} \cchi{2}{2}- q^{-2} \cchi{2}{2} \cchi{2}{1}=
-q^{-1}\cchi{2}{1},~~~~~~~~\mbox{and~same~with~barred~indices}\\
{}~~\\
\cchi{1}{0}\cchi{1}{1}-q^{-2}\cchi{1}{1}\cchi{1}{0}-
  (q^{-2}-1)\cchi{1}{2} \cchi{2}{0}=-\qm \cchi{1}{0} \\
\cchi{1}{0}\cchi{1}{2}-q^{-1}\cchi{1}{2}\cchi{1}{0}=0\\
\cchi{1}{0}\cchi{2}{1}-q^{-1}\cchi{2}{1}\cchi{1}{0}-
  (\qm-q)\cchi{2}{2} \cchi{2}{0}=-\cchi{2}{0} \\
\cchi{1}{0}\cchi{2}{2} - \cchi{2}{2}\cchi{1}{0}=0 \\
\cchi{2}{0}\cchi{1}{1}-\cchi{1}{1}\cchi{2}{0}-
  (q^{-2}-1)\cchi{2}{1} \cchi{1}{0}-(-2+q^{-2}+q^2)
  \cchi{2}{2} \cchi{2}{0}=(q-\qm) \cchi{2}{0} \\
\cchi{2}{0}\cchi{1}{2}-q^{-1}\cchi{1}{2}\cchi{2}{0}-
  (\qm-q)\cchi{2}{2} \cchi{1}{0}=- \cchi{1}{0} \\
\cchi{2}{0}\cchi{2}{1}-q^{-1}\cchi{2}{1}\cchi{2}{0}=0\\
\cchi{2}{0}\cchi{2}{2}-q^{-2}\cchi{2}{2}\cchi{2}{0}
  =-\qm \cchi{2}{0}\\
{}~~\\
\cchi{\bar 0}{\bar 1}\cchi{\bar 1}{\bar 1}-q^2 \cchi{\bar 1}{\bar 1}
\cchi{\bar 0}{\bar 1}-
  (1-2q^2+q^4)\cchi{\bar 2}{\bar 2} \cchi{\bar 0}{\bar 1}-
(-q+q^3) \cchi{\bar 2}{\bar 1}\cchi{\bar 0}{\bar 2}=q^3 \cchi{\bar 0}
{\bar 1} \\
\cchi{\bar 0}{\bar 1}\cchi{\bar 1}{\bar 2}-q\cchi{\bar 1}{\bar 2}
\cchi{\bar 0}{\bar 1}-(-q^2+q^4)\cchi{\bar 2}{\bar 2} \cchi{\bar 0}{\bar
2}=q^3 \cchi{\bar 0}{\bar 2}\\
\cchi{\bar 0}{\bar 2}\cchi{\bar 1}{\bar 1}-\cchi{\bar 1}{\bar 1}
\cchi{\bar 0}{\bar 2}-
  (q-\qm)\cchi{\bar 1}{\bar 2} \cchi{\bar 0}{\bar 1}=0 \\
\cchi{\bar 0}{\bar 2}\cchi{\bar 1}{\bar 2} - q \cchi{\bar 1}{\bar 2}
\cchi{\bar 0}{\bar 2}=0 \\
\cchi{\bar 0}{\bar 1}\cchi{\bar 2}{\bar 1} - q \cchi{\bar 2}{\bar 1}
\cchi{\bar 0}{\bar 1}=0 \\
\cchi{\bar 0}{\bar 1}\cchi{\bar 2}{\bar 2} - \cchi{\bar 2}{\bar 2}
\cchi{\bar 0}{\bar 1}=0 \\
\cchi{\bar 0}{\bar 2}\cchi{\bar 2}{\bar 1}-q\cchi{\bar 2}{\bar 1}
\cchi{\bar 0}{\bar 2}-(q^2-1)\cchi{\bar 2}{\bar 2} \cchi{\bar 0}
{\bar 1}=q \cchi{\bar 0}{\bar 1}\\
\cchi{\bar 0}{\bar 2}\cchi{\bar 2}{\bar 2}-q^2\cchi{\bar 2}{\bar 2}
\cchi{\bar 0}{\bar 2}=q\cchi{\bar 0}{\bar 2} \\
{}~~\\
\cchi{1}{0} \cchi{2}{0}-q \cchi{2}{0} \cchi{1}{0}=0\\
\cchi{\bar 0}{\bar 1} \cchi{\bar 0}{\bar 2} - \qm \cchi{\bar 0}{\bar 2}
\cchi{\bar 0}{\bar 1}=0\\
\cchi{1}{0}\cchi{\bar 0}{\bar 1}-q^{-2} \cchi{\bar 0}{\bar 1} \cchi{1}{
0}+(1-q^{-2}) \cchi{\bar 0}{\bar 2}\cchi{2}{0}=0\\
\cchi{1}{0}\cchi{\bar 0}{\bar 2}-q^{-1} \cchi{\bar 0}{\bar 2}
\cchi{1}{0}=0\\
\cchi{2}{0}\cchi{\bar 0}{\bar 1}-q^{-1} \cchi{\bar 0}{\bar 1}
\cchi{2}{0}=0\\
\cchi{2}{0}\cchi{\bar 0}{\bar 2}-q^{-2} \cchi{\bar 0}{\bar 2}
\cchi{2}{0}=0\\
\end{array}
\]
\[
\begin{array}{l}
\cchi{1}{1} \cchi{\bar 1}{\bar 1} - \cchi{\bar 1}{\bar 1}\cchi{1}{1}+
(-q^{-4}+q^{-2})\cchi{\bar 0}{\bar 1} \cchi{1}{0} +(-1-q^{-4}+2q^{-2})
\cchi{\bar 0}{\bar 2} \cchi{2}{0} -\\
{}~~~~-(1-q^2)\cchi{\bar 1}{\bar 2} \cchi{2
}{1}=0\\
\cchi{1}{1} \cchi{\bar 1}{\bar 2}- \cchi{\bar 1}{\bar 2}\cchi{1}{1}-
(-q^{-3} +q^{-1})\cchi{\bar 0}{\bar 2} \cchi{1}{0}=0\\
\cchi{1}{1} \cchi{\bar 2}{\bar 1}-\cchi{\bar 2}{\bar 1} \cchi{1}{1}-
(q^2-1)\cchi{\bar 1}{\bar 1} \cchi{2}{1} - (1-q^2)
\cchi{\bar 2}{\bar 2} \cchi{2}{1}=0\\
\cchi{1}{1} \cchi{\bar 2}{\bar 2}-\cchi{\bar 2}{\bar 2} \cchi{1}{1}
-(1-q^{-2}) \cchi{\bar 1}{\bar 2} \cchi{2}{1}=0\\
\cchi{1}{2} \cchi{\bar 1}{\bar 1} - (q^2-1) \cchi{\bar 1}{\bar 2}
\cchi{1}{1} - \cchi{\bar 1}{\bar 1} \cchi{1}{2} - (1-q^2) \cchi{\bar 1
}{\bar 2} \cchi{2}{2}=0\\
\cchi{1}{2} \cchi{\bar 1}{\bar 2} - q^2 \cchi{\bar 1}{\bar 2} \cchi{1}{2
} = 0\\
\cchi{1}{2} \cchi{\bar 2}{\bar 1} -q^{-2}
\cchi{\bar 2}{\bar 1} \cchi{1}{2}
+(-q^{-6}+q^{-4})\cchi{\bar 0}{\bar 1} \cchi{1}{0} - q^{-6}(1-q^2)^2
\cchi{\bar 0}{\bar 2} \cchi{2}{0} - \\
{}~~~~-(-1+q^{-2})\cchi{\bar 1}{\bar 1} \cchi{1}{1} -
(1-q^{-2}) \cchi{\bar 2}{\bar 2} \cchi{1}{1}-
(q^{-4}-q^{-2} + q^2-1) \cchi{\bar 1}{\bar 2} \cchi{2}{1}-\\
{}~~~~-(1-q^{-2}) \cchi{\bar 1}{\bar 1} \cchi{2}{2} -
(q^{-2} -1) \cchi{\bar 2}
{\bar 2} \cchi{2}{2}=0\\
\cchi{1}{2} \cchi{\bar 2}{\bar 2} - \cchi{\bar 2}{\bar 2} \cchi{1}{2}
+(-q^{-5}+q^{-3}) \cchi{\bar 0}{\bar 2} \cchi{1}{0} - (-1 +q^{-2})
\cchi{\bar 1}{\bar 2} \cchi{1}{1} -\\
{}~~~~-(1-q^{-2}) \cchi{\bar 1}{\bar 2}
\cchi{2}{2}=0\\
\cchi{2}{1} \cchi{\bar 1}{\bar 1} - \cchi{\bar 1}{\bar 1} \cchi{2}{1}+
(-q^{-3}+q^{-1}) \cchi{\bar 0}{\bar 1} \cchi{2}{0}=0\\
\cchi{2}{1} \cchi{\bar 1}{\bar 2} - q^{-2}\cchi{\bar 1}{\bar 2}
\cchi{2}{1}+
(-q^{-4}+q^{-2}) \cchi{\bar 0}{\bar 2} \cchi{2}{0}=0\\
\cchi{2}{1} \cchi{\bar 2}{\bar 1} - q^{2}\cchi{\bar 2}{\bar 1}
\cchi{2}{1}=0\\
\cchi{2}{1} \cchi{\bar 2}{\bar 2} - \cchi{\bar 2}{\bar 2}
\cchi{2}{1}=0\\
\cchi{2}{2} \cchi{\bar 1}{\bar 1} - \cchi{\bar 1}{\bar 1}
\cchi{2}{2}-
(1-q^{-2}) \cchi{\bar 1}{\bar 2} \cchi{2}{1}=0\\
\cchi{2}{2} \cchi{\bar 1}{\bar 2} - \cchi{\bar 1}{\bar 2}
\cchi{2}{2}=0\\
\cchi{2}{2} \cchi{\bar 2}{\bar 1} - \cchi{\bar 2}{\bar 1}
\cchi{2}{2}+(-q^{-5}+q^{-3})\cchi{\bar 0}{\bar 1} \cchi{2}{0} - (-1+q^{-
2}) \cchi{\bar 1}{\bar 1} \cchi{2}{1} -\\
{}~~~~-(1-q^{-2}) \cchi{\bar 2}{\bar 2}
\cchi{2}{1}=0\\
\cchi{2}{2} \cchi{\bar 2}{\bar 2} - \cchi{\bar 2}{\bar 2}
\cchi{2}{2}+(-q^{-6}+q^{-4})\cchi{\bar 0}{\bar 2} \cchi{2}{0} - (q^{-4}-
q^{-2}) \cchi{\bar 1}{\bar 2} \cchi{2}{1}=0\\
{}~~\\
\cchi{1}{1} \cchi{\bar 0}{\bar 1} - \cchi{\bar 0}{\bar 1} \cchi{1}{1}
-(1-q^2) \cchi{\bar 0}{\bar 2} \cchi{2}{1} =0\\
\cchi{1}{1} \cchi{\bar 0}{\bar 2} - \cchi{\bar 0}{\bar 2} \cchi{1}{1}
=0\\
\cchi{1}{2} \cchi{\bar 0}{\bar 1} - q^{-1} \cchi{\bar 0}{\bar 1}
\cchi{1}{2}
-(q-q^{-1}) \cchi{\bar 0}{\bar 2} \cchi{1}{1} -(\qm-q) \cchi{\bar 0
}{\bar 2} \cchi{2}{2} =0\\
\cchi{1}{2} \cchi{\bar 0}{\bar 2} - q\cchi{\bar 0}{\bar 2} \cchi{1}{2}
=0\\
\cchi{2}{1} \cchi{\bar 0}{\bar 1} - q\cchi{\bar 0}{\bar 1} \cchi{2}{1}
=0\\
\cchi{2}{1} \cchi{\bar 0}{\bar 2} - \qm\cchi{\bar 0}{\bar 2} \cchi{2}{1}
=0\\
\cchi{2}{2} \cchi{\bar 0}{\bar 1} - \cchi{\bar 0}{\bar 1} \cchi{2}{2}
-(1-q^{-2}) \cchi{\bar 0}{\bar 2} \cchi{2}{1}=0\\
\cchi{2}{2} \cchi{\bar 0}{\bar 2} - \cchi{\bar 0}{\bar 2} \cchi{2}{2}
=0\\
{}~~\\
\cchi{1}{0} \cchi{\bar 1}{\bar 1} - \cchi{\bar 1}{\bar 1} \cchi{1}{0}
-(1-q^2) \cchi{\bar 1}{\bar 2} \cchi{2}{0}=0\\
\cchi{1}{0} \cchi{\bar 1}{\bar 2} - q\cchi{\bar 1}{\bar 2} \cchi{1}{0}
=0\\
\cchi{1}{0} \cchi{\bar 2}{\bar 1} - \qm\cchi{\bar 2}{\bar 1} \cchi{1}{0}
-(q-\qm) \cchi{\bar 1}{\bar 1} \cchi{2}{0}-(\qm-q)\cchi{\bar 2}{\bar 2}
\cchi{2}{0}=0\\
\cchi{1}{0} \cchi{\bar 2}{\bar 2} - \cchi{\bar 2}{\bar 2} \cchi{1}{0}
-(1-q^{-2}) \cchi{\bar 1}{\bar 2} \cchi{2}{0}=0\\
\cchi{2}{0} \cchi{\bar 1}{\bar 1} - \cchi{\bar 1}{\bar 1} \cchi{2}{0}
=0\\
\cchi{2}{0} \cchi{\bar 1}{\bar 2} - \qm \cchi{\bar 1}{\bar 2} \cchi{2}{0}
=0\\
\cchi{2}{0} \cchi{\bar 2}{\bar 1} - q \cchi{\bar 2}{\bar 1} \cchi{2}{0}
=0\\
\cchi{2}{0} \cchi{\bar 2}{\bar 2} - \cchi{\bar 2}{\bar 2} \cchi{2}{0}
=0\\
\end{array}
\]

\vfill\eject

\vfill\eject

\begin{thebibliography}{99}

\bibitem{Drinfeld} V. Drinfeld, Sov. Math. Dokl. {\bf 32} (1985) 254.

\bibitem{Jimbo} M. Jimbo, Lett. Math. Phys. {\bf 10} (1985) 63; {\bf 11}
        (1986) 247.

\bibitem{FRT} L.D. Faddeev,
N.Yu. Reshetikhin and L.A. Takhtajan, Algebra and
Analysis, {\bf 1} (1987) 178.

\bibitem{Majid1} S. Majid, Int. J. Mod. Phys. {\bf A5} (1990) 1.

\bibitem{Wor} S.L. Woronowicz, Publ. RIMS, Kyoto Univ., Vol. {\bf 23}, 1
(1987) 117;  Commun. Math. Phys. {\bf 111} (1987) 613 and
Commun. Math. Phys. {\bf 122}, (1989) 125.

\bibitem{Bernard} D. Bernard, {\sl Quantum Lie
algebras and differential calculus on quantum groups}, Proc. 1990 Yukawa
Int. Seminar, Kyoto; Phys. Lett. {\bf 260B} (1991) 389.

\bibitem{Jurco} B. Jur\v{c}o, Lett. Math. Phys. {\bf 22} (1991) 177.

\bibitem{Zumino} B. Zumino,
{\sl Introduction to the Differential Geometry
of Quantum Groups}, LBL-31432 and UCB-PTH-62/91, notes
of a plenary talk given at the 10-th IAMP Conf., Leipzig (1991);
P. Schupp, P. Watts and B. Zumino, {\sl Differential
Geometry on Linear Quantum Groups}, preprint LBL-32314, UCB-PTH-92/13
(1992).

\bibitem{Aschieri} P. Aschieri and
L. Castellani, {\sl An
introduction to non-commutative
differential geometry on quantum groups}, preprint
CERN-TH.6565/92, DFTT-22/92 (1992), to be publ. in Int. Jou.
Mod. Phys. A.

\bibitem{Cas1} L. Castellani, {\sl R matrix and bicovariant calculus
for the inhomogeneous quantum groups $IGL_q(n)$}, Torino preprint
DFTT-59/92, to be publ. in Phys. Lett. {\bf B}.

\bibitem{Schlieker} M. Schlieker, W. Weich and R. Weixler, {\sl
Inhomogeneous quantum groups}, LMU-TPW 1191-3.

\bibitem{Drabant} B. Drabant, M. Schlieker, W. Weich and B. Zumino,
{\sl Complex quantum groups and their quantum enveloping algebras},
MPI-PTh/91-75 and LMU-TPW 1991-5.

\bibitem{qLorentz} P. Podle\'s and
S.L. Woronowicz, Commun. Math. Phys. {\bf 130
} (1990) 381;
U. Carow-Watamura, M. Schlieker, M. Scholl and
S. Watamura, Z. Phys. {\bf
C48} (1990) 159 and Int. Jou. Mod. Phys. {\bf A6} (1991) 3081;
W.B. Schmidke, J. Wess and B. Zumino, Z. Phys. {\bf C52} (1991) 471;
O. Ogievetsky, W.B. Schmidke, J. Wess and B. Zumino, MPI-PTh/91-51
(1991);
S. Majid, DAMPT/92-10;
X.D. Sun, S.K. Wang and K. Wu, CCAST-92-14, ASIAM-92-11,
ASITP-92-18 (1992).

\bibitem{qPoincare}J. Lukierski, A. Novicki, H. Ruegg and V.N. Tolstoy,
Phys. Lett. {\bf B264} (1991) 331; J. Lukierski, A. Novicki and
H. Ruegg, Phys. Lett. {\bf B271} (1991) 321 and {\bf B293}
(1992) 344; V. Dobrev, in the
Proceedings of the Quantum Groups Workshop, Goslar, 1991;
O. Ogievetsky, W.B. Schmidke, J. Wess and
B. Zumino, MPI-Ph/91-98, LBL-31703, UCB 92/04;
S. Majid, DAMTP/92-65.

\bibitem{Aschieri2} P. Aschieri and L. Castellani, Phys. Lett.
{\bf B293} (1992) 299.

\bibitem{Cas2} L. Castellani, {\sl Bicovariant differential calculus on
the quantum Poincar\'e group}, Torino preprint DFTT-66/92


\end{thebibliography}
\end{document}